# Stratospheric balloons as a platform for the next large far infrared observatory

P. Maier[a]\*, J. Wolf[a], A. Krabbe[a], T. Keilig[a], A. Pahler[a], S. Bougueroua[a], T. Müller[b], R. Duffard[c], J.-L. Ortiz[c], S. Klinkner[a], M. Lengowski[a], C. Krokstedt[d], C. Lockowandt[d], N. Kappelmann[e], B. Stelzer[e], K. Werner[e], S. Geier[e,f], C. Kalkuhl[e], T. Rauch[e], T. Schanz[e], J. Barnstedt[e], L. Conti[e], L. Hanke[e], M. Kaźmierczak-Barthel[a]

[a] *Institute of Space Systems, University of Stuttgart, Pfaffenwaldring 29, 70569 Stuttgart*, Germany, pmaier@irs.uni-stuttgart.de
[b] *Max-Planck-Institut für extraterrestrische Physik, Giessenbachstraße, 85741 Garching, Germany*
[c] *Instituto de Astrofísica de Andalucía, Glorieta de la Astronomía S/N, 18008 Granada, Spain*
[d] *Swedish Space Corporation, Torggatan 15, 17154 Solna, Sweden*
[e] *Institut für Astronomie und Astrophysik, Universität Tübingen, Sand 1, 72076 Tübingen, Germany*
[f] *Institut für Physik und Astronomie, Universität Potsdam, Karl-Liebknecht-Str. 24/25, 14467 Potsdam, Germany*
\* Corresponding Author

**Abstract**

Observations that require large physical instrument dimensions and/or a considerable amount of cryogens, as it is the case for high spatial resolution far infrared (FIR) astronomy, currently still face technological limits for their execution from space. Due to these limits, the FIR domain in particular is lagging behind other wavelength regimes in terms of angular resolution and available observational capabilities, especially after the retirement of the Herschel Space Observatory.

Balloon-based platforms promise to complement the existing observational capabilities by offering means to deploy comparably large telescopes with comparably little effort, including other advantages such as the possibility to regularly refill cryogens and to change and/or update instruments.

The planned European Stratospheric Balloon Observatory (ESBO), currently under preparation by a consortium of European research institutes and industry, aims at providing these additional large aperture FIR capabilities, exceeding the spatial resolution of Herschel, in the long term. In particular, the plans focus on reusable platforms performing regular flights and an operations concept that provides researchers with proposal-based access to observations as also practiced on space-based observatories. It thereby aims at offering a complement to other airborne, ground-based and space-based observatories in terms of access to wavelength regions, spatial resolution capability, and photometric stability. In order to fully exploit the potential offered by regularly flying balloon platforms, ESBO foresees the option to exchange instruments and telescopes in between flights.

While the FIR capabilities are a main long-term objective, ESBO will offer benefits in other wavelength regimes along the way. Within the recently initiated ESBO *Design Study* (ESBO *DS*), financed within the European Union's Horizon 2020 Programme, a prototype platform carrying a 0.5 m telescope for ultraviolet (UV) and visible (VIS) light observations is being built and a platform concept for a next-generation FIR telescope is being studied. A flight of the UV/VIS prototype platform is currently foreseen for 2021.

In this paper we will outline the scientific and technical motivation for a large aperture balloon-based FIR observatory and the ESBO *DS* approach towards such an infrastructure.

Secondly, we will present the technical motivation, science case, and instrumentation of the 0.5 m UV/VIS platform.

**Acronyms/Abbreviations**

| | |
|---|---|
| ALMA | Atacama Large Millimeter/submillimeter Array |
| BEXUS | Balloon Experiments for University Students |
| BLAST | Balloon-borne Large Aperture Submillimeter Telescope |
| DLR | Deutsches Zentrum für Luft- und Raumfahrt |
| ESA | European Space Agency |
| ESBO *DS* | European Stratospheric Balloon Observatory *Design Study* |
| ESO | European Southern Observatory |
| FIR | Far Infrared |
| IAAT | Institut für Astronomie und Astrophysik Tübingen |
| IRAS | Infrared Astronomical Satellite |
| JWST | James Webb Space Telescope |
| MCP | Micro-Channel Plate |
| NIR | Near Infrared |
| NOEMA | Northern Extended Millimeter Array |
| ORFEUS | Orbiting and Retrievable Far and Extreme Ultraviolet Spectrometer |
| ORISON | innOvative Research Infrastructure |
| | based on Stratospheric balloONs |
| PILOT | Polarized Instrument for the Long-wavelength Observation of the Tenuous ISM |
| SOFIA | Stratospheric Observatory For Infrared Astronomy |
| SPICA | Space Infrared Telescope for Cosmology and Astrophysics |
| SSC | Swedish Space Corporation |
| STO | Stratospheric Terahertz Observatory |
| STUDIO | Stratospheric UV Demonstrator of an Imaging Observatory |
| UV | Ultraviolet |
| THISBE | Telescope of Heidelberg for Infrared Studies by Balloon-Borne Experiments |
| VIS | Visible |
| WD | White Dwarf |

## 1. Introduction

*1.1 The current situation of balloon-based telescopes*

The idea of using stratospheric balloons to overcome the obstructions of Earth's atmosphere for astronomical observations is not new. Historically, the advantages were obvious: spacecraft did not exist and capabilities of planes were limited, leaving balloons as the only option to move instruments above most of the atmosphere. In current times, the benefits do not seem as clear: both spacecraft and planes provide powerful observation platforms and ground-based telescopes invest large efforts into compensating atmospheric influences. However, even in the era of nano- and microsatellites, space observatories are intrinsically expensive and bear operational limitations: development times are long, updates or corrections of the instrumentation are usually not possible after launch, operating material such as cryogenic coolant fluids cannot be refilled or replaced (see the Herschel Space Observatory). Furthermore, comparably conservative approaches towards new technologies are used to minimize risks of expensive failure. Ground-based and airborne telescopes, on the other hand, still suffer from fundamental limitations imposed by the atmosphere at certain wavelengths.

Nevertheless, the use of balloon-based telescopes is limited. With some notable exceptions (the Telescope of Heidelberg for Infrared Studies by Balloon-Borne Experiments (THISBE) [1], the Balloon-borne Large Aperture Submillimeter Telescope (BLAST) [2]), modern balloon-based telescopes have mostly served a single purpose and did not fly more than a few times. It seems likely that the main reasons are a combination of the challenges associated with large balloon payloads (such as safe payload recovery and precise pointing) and the specialized expertise required for large stratospheric balloon missions that most astronomical research groups do not have.

The goal of ESBO is to systematically address the technical challenges from an astronomical observatory's point of view and to lower the entry barrier to balloon-based observations by providing an operating institution that offers observing time and instrument space on balloon-based telescopes.

*1.2 ESBO Project History and Future*

Over the last two years, the H2020-funded project ORISON (innOvative Research Infrastructure based on Stratospheric balloONs) assessed interest and scientific needs within the (mostly European) astronomical community with regard to balloon-based research infrastructures and studied the general feasibility of a balloon-based observatory [3].

The plans for ESBO pick up from the positive conclusions of ORISON and aim at creating an observatory institution based on the following cornerstones:

- Provision of instrument flight opportunities as well as open observation time access to the scientific community,
- Operation of regularly flying balloon telescopes and provision of related services by an operating institution,
- Provision of the regular opportunity to refill consumables, upgrade and/or exchange instruments in between flights,
- Maximum reuse of platform hardware in between flights to ensure efficient and fast turnaround in between flights, which includes safe recovery of hardware.

More details on the envisioned ESBO infrastructure and its scope can be found in section 5 of this paper.

The ongoing ESBO *Design Study* represents the second step towards ESBO. Under ESBO *DS*, the full infrastructure, with particular technical focus on FIR observational capabilities, is being conceptually designed. In addition, a prototype UV/visible flight system (the Stratospheric UV Demonstrator of an Imaging Observatory, STUDIO) is being developed and built to test some of the key technologies identified.

The focus of this paper is on the capabilities that an ESBO platform could offer for the FIR. It will thus firstly lay out the scientific and technical motivation for a new FIR telescope in section 2. In section 3 the advantages and potential performance of a balloon-based FIR telescope will be described. Finally, the UV/VIS prototype currently under development as well as an outlook onto a potential development timeline will be presented shortly in sections 4 and 5.

## 2. Motivation for a far infrared telescope

*2.1 Current situation*

As the FIR range is not accessible from the ground, the FIR astronomy community only slowly started to develop (and is still developing) thanks to a series of space-based and airborne observatories. Most of them, however, have been spacecraft with a limited lifetime or instruments with limited spectral range or resolution. As such, after the end of the Herschel mission, the community is currently left with only one active FIR observatory, the Stratospheric Observatory For Infrared Astronomy (SOFIA), and sporadically flying balloon missions (such as BLAST [4], the Stratospheric Terahertz Observatory (STO) [5], or the Polarized Instrument for the Long-wavelength Observation of the Tenuous ISM (PILOT) [6], see also Table 1).

Table 1. Active far infrared observatories.

| Mission | Wavelength coverage | Effective aperture diam. |
|---|---|---|
| SOFIA[*] | 0.36 – 612 µm | 2.5 m |
| BLAST [10] | 240, 350, 500 µm | 2.5 m |
| PILOT [6] | 240 & 550 µm | 1 m |
| STO [5] | 158, 205 µm | 0.8 m |

Astronomers, especially astrochemists, are still waiting for new FIR telescopes. It is thus the time to plan the next mission that will cover the gap in the FIR sky between JWST's upper limit (28.5 µm) and ALMA's lower limit (316 µm, Band 10).

Next steps of FIR science will further investigate the origins of water on planets in our and distant solar systems, study mechanisms and details of star and planet formation by investigating chemical evolution and cooling processes throughout the universe, and further investigate the Interstellar Medium, its interaction with stellar environments, and its energy cycle, by observations of dust and gas (as expressed e.g. by the European Far-Infrared Space Roadmap [7]).

Taking these scientific steps forward requires telescopes with better angular resolution, more observational capacity (in terms of spectral coverage and observation time), and higher sensitivity. The latter need would be addressed by the Space Infrared Telescope for Cosmology and Astrophysics (SPICA, 2.5 m telescope diameter) [8], which was recently downselected as one of the three finalists for the ESA M5 mission. The first need, however, can only be achieved by larger telescopes or interferometric observations. In addition, large telescope apertures are required in order to overcome the so-called confusion limit, the notion that strong signals from dust and gas in the foreground make small astronomical targets in their background indistinguishable. Logistical and technical challenges, however, make space-based telescopes with large apertures, as required for high angular resolution observations, extremely difficult. Balloon-borne telescopes do not face many of these challenges and thus are particularly well suited to address the first two needs, while offering the possibility to regularly use the most up-to-date instrumentation. Such a concept has already been proven successful by SOFIA.

*2.2 Scientific motivation*

As this paper focuses on a potential observatory for the FIR, several potential key science areas presented in the following are grouped by the sort of observation they require.

*2.2.1 Discrete sources*

Particularly the advancement of research concerning star and planet formation and astrochemistry, but also solar system science, relies upon further information from FIR observations. Outstanding topics in these areas include the further study of cold dust and ices, that of light hydrides, and the study of the distribution of molecules in general, be it atmospheres of solar system objects, the Milky Way, or other galaxies. In the following, we will highlight a few of the prominent science case whose study a large aperture balloon observatory would enable.

Ice features in the FIR

Dust, for years, annoyed astronomers by covering their favourite stars as well as the birth places of those stars and their planetary systems. With the development of infrared observatories, however, cold dust and ices became a hot topic, allowing important insight into the process of star and planet formation and into the migration process of water through evolving planetary systems. So far, mainly features in the near- and mid-infrared have been used to detect and characterize molecular ices in dark clouds and protoplanetary disks. As their FIR emission features are attributed to intermolecular vibration modes, however, observing them in the FIR makes it furthermore possible to determine the structure and transitions between phases of the observed medium (e.g. amorphous vs. crystalline). In particular, the FIR band positions and widths are, in addition to the abundance of the emitting species, sensitive to the grain geometry and size distribution, the environment temperature and density structure. Combined with modelling of protoplanetary disk emissions, the analysis of the FIR features thus allows to infer the abundance and location of ices within the disk, making it, with sufficient data, possible to constrain the location of the snow line (the distance from a star/protostar where it is cold enough for volatiles to condense into ice) [11].

---

[*] Current coverage in Observing Cycle 7, wavelength coverage partly interrupted by atmospheric extinction [9].

So far, only water ice features have been detected in the FIR in a few disks, while the band strengths of other ices are thought to be not strong enough to have been detected. A sensitive balloon observatory offering medium spectral resolution observations at the wavelengths of these ices would help to study them in many targets across the Galaxy (for band locations and band strengths of prominent molecular ices from laboratory measurements see Giuliano et al. [12]).

Light hydrides

Light hydrides, on the other hand, belong to the first molecules to form in atomic gas and are thus at the starting point of astrochemistry and the building blocks of larger molecules [13]. Their study allows fundamental insight into the first building steps towards interstellar molecules. As their chemical formation process only involves a few steps, the interpretation of their abundances is comparably straightforward and they can provide key information about their environments, including on dynamical processes (shocks, turbulence, large scale winds), cosmic ray ionization rate, and presence of molecular hydrogen [13].

The observation of light hydrides thus promises to provide a valuable tool to understand planet and star formation, and, through the measurement of isotopic ratios, also to understand the origin of volatiles in our own solar system. While the idea behind the study of light hydrides sounds relatively simple, their observations are much more complex. They require very high spectral resolution mostly in wavelength ranges that are not accessible at lower altitudes. ALMA and NOEMA are already enabling highly sensitive and spatially highly resolved observations of light hydrides in distant galaxies, for which the ground state transition lines are sufficiently redshifted to fall into the sub-mm/mm spectral regions. For observations in the interstellar medium in our own Galaxy, or neighbouring galaxies, however, the lines have to be observed at (or close to) their original wavelengths in the FIR. A sensitive balloon observatory offering high spectral resolution observations at the wavelengths of light hydrides ground states would thus allow their study in many targets across our Galaxy.

Solar System atmospheres

In the solar system context, balloon-based observations would allow e.g. studies of the vertical distributions of molecules in atmospheres of planets, their satellites, or in comae. These observations require high spectral resolution to determine the shape of absorption lines. While many molecules could theoretically be observed with SOFIA, telluric absorption lines are still considerably pressure broadened in the remaining atmosphere. At 30 to 40 km altitude, the line width of telluric absorption lines is narrow enough to allow to distinguish between the telluric lines and Doppler shifted absorption features on solar system objects.

*2.2.2 Surveys*

The FIR spectral range is generally lagging behind with regard to high-resolution large-scale maps of our Galaxy, as compared to the adjacent sub-mm and mid infrared spectral regions. The only complete continuum sky survey at 100 µm existing today, for example, is the one taken by the Infrared Astronomical Satellite (IRAS) with a resolution of 1.5 arcminutes. The situation is similar with regard to dedicated spectral line surveys. Particularly the 157.7 µm (CII) line and the 63.18 µm (OI) line may radiate up to several percent of the entire Galaxy's energy output, and Herschel and SOFIA have been able to study these important lines a bit already. However, very little is still known about their spatial distribution across our Galaxy. Without capacities in the FIR, this situation will not change, as Figure 1 indicates. Mapping out these and other important FIR spectral lines (such as the 128 µm HD line) across a major fraction of our own Galaxy as well as of other galaxies with a high signal to noise ratio will boost our understanding of the chemical evolution of our Galaxy and of galactic evolution in general.

A balloon observatory can achieve about 1000 hours of observing time during a 6-weeks mission with one instrument attached. Such a set-up is very much suited for executing large surveys.

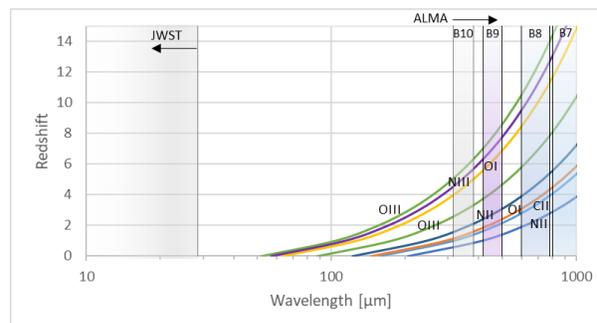

Fig. 1. Spectral locations of major atomic lines, including the OI and CII cooling lines at different redshifts. At the left and right ends of the figure, the spectral ranges covered by JWST and the shorter-wavelength ALMA bands (10, 9, 8, and 7) are marked. As the figure indicates, in our own Galaxy these lines can only be observed in the far infrared.

*2.3 Technical Motivation*

Due to the fundamental inaccessibility of the FIR spectrum from the ground, FIR observations are solely reliant upon air-, balloon-, or space-borne capabilities. Each of these observational platforms comes with its own strong and weak points. When it comes to the abovementioned need for high sensitivity, this can likely best be addressed by space telescopes with cooled optics, such as SPICA. High angular resolution, however, requires either large mirrors or interferometric observations. Large mirrors pose significant challenges to space missions which, in order to overcome the size limitation imposed by the launcher fairings, have to rely upon complex and expensive unfolding mechanisms such as used on the JWST [14]. High vibrational loads during launch furthermore complicate the application of very large optical structures in space. In order to implement space-based interferometric observations, which would offer an alternative to large single-dish telescopes, other serious technical challenges still remain to be solved.

Similarly, the 2.7 m telescope on SOFIA is about the largest telescope size that can be accommodated on an existing plane that reaches the required altitude. The Airbus A380, that may seem like a possible alternative, only has a slightly larger fuselage crosssection than SOFIA's Boeing 747SP, and a flight ceiling of only 13.1 km as compared to 13.7 km of SOFIA (the Airbus A300-600ST "Beluga" has a larger useable crosssection, but only a flight ceiling of approx. 11 km).

Platforms suspended underneath stratospheric balloons, however, do not encounter these size constraints. They also do not require large telescope structures to withstand the high vibrational and shock loads as experienced during rocket launches.

In addition, balloon flights with regular touch-down allow the use of consumable cryogens and, particularly important for survey missions, practically do not constrain the amount of data to be obtained, as long as the major share of it can be retrieved at the end of a mission.

## 3. Advantages of a balloon-based FIR observatory
### 3.1 Potential implementation of an ESBO FIR platform
#### 3.1.1 Telescope sizing

As indicated in the motivation, two of the main driving needs for a balloon-based FIR platform are the improvement on angular resolution and the overcoming of the current confusion limit. Both needs drive the aperture size of the telescope. To provide significant improvement over current capabilities, we propose an observatory with a 5 m aperture, improving on the current SOFIA resolution by a factor of 2 and providing 100 µm observations at 5 arcsec, corresponding to the resolution of the Herschel surveys at 70 µm (see also Figure 2).

To keep the telescope size and mass manageable, an architecture option would be using an off-axis Gregorian telescope layout with a 5 m x 2.5 m elliptical primary mirror. With state of the art materials, we would expect such a mirror to be of similar mass as the Herschel primary. Improvements with regard to carbon composite mirrors, however, may make an areal density of 20 kg/m$^2$ possible, which would decrease the mass of the primary to around 200 kg.

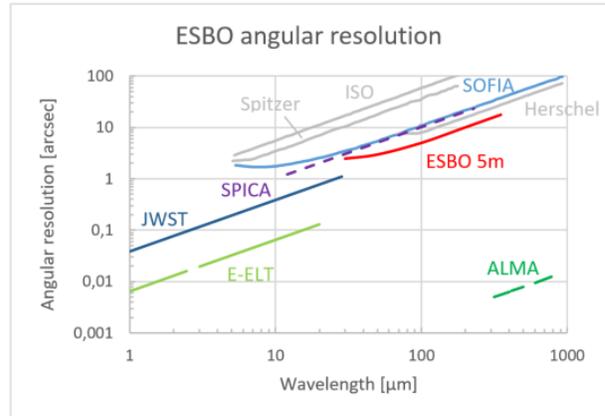

Fig. 2. Comparison of angular resolution of a 5 m aperture balloon telescope to other past, existing, and planned observatories. Past missions: ISO, Spitzer, Herschel; existing: SOFIA, ALMA; in preparation: JWST, E-ELT; proposed: SPICA.

#### 3.1.2 Flight altitude and options

Large stratospheric balloons can nowadays readily achieve flight altitudes of 40 km with several tons of payload. As Figure 3 indicates, however, for a FIR observatory, a moderate flight altitude of 30 km already would provide close to no atmospheric extinction as well as stable conditions. For this altitude, regular zero-pressure stratospheric balloons provide payload capacities of up to 3600 kg [15] and flight durations of more than 40 days, using circumpolar flight routes over Antarctica. Flown once a year, such flights would offer about the same total yearly observation time as SOFIA.

The current progress in the development of super pressure balloons may, however, even promise flight times of up to 100 days from mid-latitude launch sites within the next couple of years [15]. In addition, other options for flight logistics than using the traditional zero or super pressure balloon flights are still under investigation.

#### 3.1.3 Operational aspects

The unique operating conditions of balloons, in addition to the observational advantages, provide some considerable advantages with regard to scientific operation, particularly compared to space missions:
- Change between different instruments in between flights;
- Update or refurbishment of instruments in between flights;
- Refill of consumables (e.g. cryogens) in between flights;

- Availability of high amounts of on-board data storage retrievable after flights, making it possible to collect more data and saving on effort for on-board data compression.

### 3.2 Observational performance

The most important advantage, however, are the excellent, space-like observation conditions in the high stratosphere. Figure 3 compares the transmission curves of FIR radiation at altitudes of different platforms (all data obtained from ATRAN simulations [16]). While the data is smoothed, it already shows that FIR observations at 30 km are practically not affected by the atmosphere.

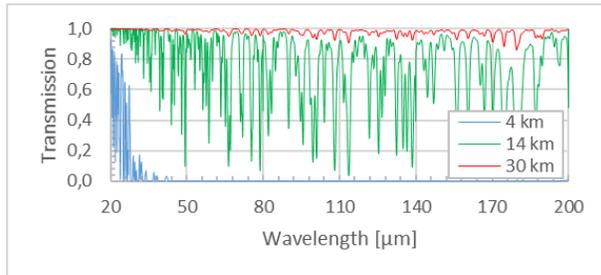

Fig. 3. Comparison of atmospheric transmission in the FIR for a high ground based observatory, at SOFIA flight altitude (14 km) and at a moderate balloon flight altitude (30 km).

While the main strength of a large balloon-based FIR telescope would be in the angular resolution rather than in sensitivity, the good observation conditions are also reflected in the expected sensitivities.

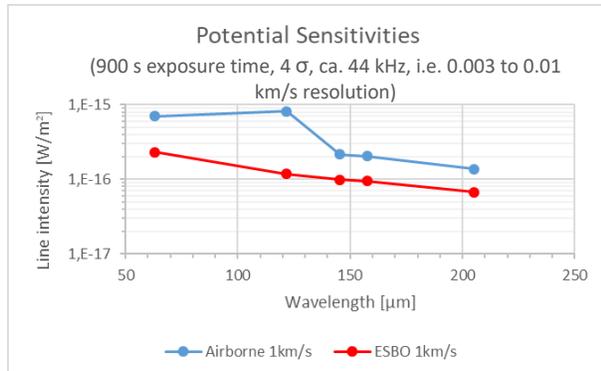

Fig. 4. Estimated minimum detectable line fluxes for single point observations of prominent atomic lines (OI, NII, OI, CII, NII) for a high spectral resolution heterodyne instrument at 30 km altitude. Values in between data points are interpolated. The plot clearly shows that the sensitivity at least doubles, which results in ¼ of the obsering time required.

For a state-of-the-art high spectral resolution heterodyne instrument, without assuming any improvements on the instrument, the sensitivity would improve greatly on the conceptual ESBO FIR platform, as Figure 4 indicates for some of the prominent FIR atomic lines.

In addition, many additional spectral regions that are not observable at 13 km become accessible at 30 km. This is particularly well illustrated by the case of light hydrides, where many lines become observable only at balloon altitudes. Table 2 lists a selection of those lines, along with sensitivity estimates for high spectral resolution (44 kHz) observations.

Table 2. Estimated minimun detectable line fluxes for high spectral resolution (ca. 44 kHz) of selected lines of light hydrides. Sensitivities are 4 σ, for 900 s exposure time and assumed line widths of 1 km/s.

| | | Line sensitivity [W/m$^2$] | |
|---|---|---|---|
| **Species** | **Wavelength** | **14 km** | **30 km** |
| $H_3O^+$ | 181.05 μm | $8.95 \cdot 10^{-16}$ | $9.36 \cdot 10^{-17}$ |
| $H_3O^+$ | 100.87 μm | - | $1.59 \cdot 10^{-16}$ |
| $H_3O^+$ | 100.58 μm | $5.82 \cdot 10^{-16}$ | $1.52 \cdot 10^{-16}$ |
| $CH^+$ | 179.62 μm | - | $9.81 \cdot 10^{-17}$ |
| $CH^+$ | 90.03 μm | - | $1.68 \cdot 10^{-16}$ |
| $CH^+$ | 72.14 μm | $5.59 \cdot 10^{-16}$ | $2.06 \cdot 10^{-16}$ |
| HF | 121.70 μm | - | $1.63 \cdot 10^{-16}$ |
| HF | 81.22 μm | - | $7.76 \cdot 10^{-16}$ |

### 3.3 Challenges and disadvantages of FIR ballooning

Despite all the advantages in observing conditions, large balloon telescopes do not come without disadvantages and technical challenges. One of them is that having only one gondola, merely one long duration flight is possible per year during Antarctic summer over Antarctica. This limits the time of year at which targets can be observed as well as the total observation time available over a year. It may change with the further development of super pressure balloons and potential alternative operating concepts (see above), but it is a standing limitation.

One of the main technical challenges is minimizing the risk of damage to equipment and achieving fast and efficient turnarounds in between flights. A potential solution to this challenge will be addressed by the STUDIO prototype, which will test, among other things, a steerable soft landing system.

## 4. UV/visible prototype platform (STUDIO)

The prototype UV/visible platform STUDIO currently under development will serve two purposes: on the one hand, it will serve as a technology demonstrator and testbed for critical technologies required for the further implementation of ESBO. On the other hand, the platform will also function as a testbed for a new microchannel plate (MCP) detector for the UV and will be available for first scientific test flights after the end of the ESBO *DS* project.

*4.1 Technical motivation*

The critical technologies to be demonstrated include a modular and scalable gondola to accommodate different astronomical payloads and their respective needs, systems for safe recovery, and a versatile, highly precise image stabilization system. For safe recovery, the foreseen systems include autonomous steered parafoils for controlled landings, in combination with a suitable gondola design, as well as a shock damping mechanism to decrease the parachute opening load. The image stabilization system will be a two-stage system, with coarse pointing control provided by the gondola and a closed-loop fine image stabilization system based on a tip-tilt mirror included within the optical system.

*4.2 Science case*

The scientific application of STUDIO makes use of the greatly reduced atmospheric extinction of UV radiation at the pursued flight altitude of 40 km. While observations below 320 nm are not possible from the ground due to extinction by ozone, they are feasible down to ~200 nm at 40 km. Only in the range 240-260 nm, residual ozone reduces the transmission to ~20%.

Two science cases making use of this spectral range motivate the scientific part of STUDIO and are shortly described in the following.

*4.2.1 Search for variable hot compact stars*

Hot and compact stars are the rather short-lived end stages of stellar evolution. They comprise the hottest white dwarfs (WDs) and hot subdwarfs. A significant fraction of them show light variations with periods ranging from seconds to hours. Among them are diverse types of pulsators, which are important to improve asteroseismic models. Others are members of ultracompact binaries (e.g., WD+WD pairs) and are strong sources of gravitational wave radiation and crucial calibrators for the future space mission eLISA. They are also regarded as good candidates for the progenitors of thermonuclear supernovae. Furthermore, compact binaries are formed via common envelope evolution and are important to study this poorly understood phase of binary evolution.

Hot compact stars have so far been studied predominantly at high galactic latitudes. Due to their very blue colours they stick out in old stellar populations like the galactic halo. However, the density of stars at high galactic latitudes is rather small and those objects are therefore very rare. Due to the 1000-times higher stellar density, the galactic disc should contain many more of those objects. Searches in the galactic plane are desirable but the identification of these faint stars is hampered by the dense, crowded fields. But not so in the UV band. The hot stars are much easier to detect there, because their emitted flux is increasing towards the UV, while the flux of the majority of other stars decreases because of their lower temperatures. Surveying the galactic plane with a UV imaging telescope will uncover many new variable hot stars.

*4.2.2 Detection of flares from cool dwarf stars*

Red dwarf stars (spectral type M) are hydrogen-burning main sequence stars like our Sun, but less massive, cooler and less luminous. The large majority of the stars in our Milky Way belongs to this group. Red dwarfs emit most of their radiation in the visible and near-infrared wavelength regions. Their UV and X-ray emission, despite being energetically a minor contribution to the overall radiation budget, ionizes material surrounding the stars and is, therefore, of central interest for the evolution of planets and other circumstellar matter. This high-energy emission of red dwarf stars is highly dynamic.

One characteristic phenomenon are flares that are stochastic brightness outbursts resulting from reconfigurations of the stellar magnetic field. During such flares, these normally faint stars become much brighter for the duration of minutes. A strong emission line of ionized magnesium (MgII) at 280 nm, covered by the STUDIO instrument, can carry up to 50% of the near-UV flux during flares. Up to now, no systematic monitoring of "flare stars" exists. Consequently, the flare occurrence rate is unknown as well as the flare energy number distribution. Particularly interesting for the study of the physics of flares is their multi-wavelength behaviour (time lags, relative energy in different bands). However, only a few simultaneous UV and optical observations exist. STUDIO enables such observations by continued monitoring (over hours or multi-epoch) of stars across the field or by focusing on prominent objects.

*4.3 Payload*

STUDIO comprises a 50 cm aperture two-mirror telescope and two simultaneously operating imaging instruments as a UV and a VIS/NIR channel. The VIS/NIR instrument will mainly be used as the sensor within a closed-loop image stabilisation system providing a pointing stability of 0.5 arcsec for the scientific observations. The main scientific instrument will be the UV instrument covering the 180 - 330 nm band at an angular resolution of approx. 1 arcsec and with a field of view of 30 arcmin x 30 arcmin. The instrument is an imaging and photon counting microchannel plate (MCP) detector, developed and built by the "Institut für Astronomie und Astrophysik Tübingen" (IAAT). It is a successor to the echelle detector whose development and flights for the ORFEUS missions are part of the space heritage of the IAAT [17]. The detector will be capable of processing about 200,000 – 300,000 detected photons per second. It is also foreseen to implement a filter wheel carrying two filters: Sloan u ($\lambda_{mean}$ = 349.8 nm, [18]) and Galex NUV ($\lambda_{mean}$ = 231.57 nm, [19]) as well as an open position. The mass of the whole detector is about 4.4 kg (including high voltage power supply and cables). The power consumption still needs to be determined but will be below 19 W (peak). A detailed description of the UV detector and its working principle can be found in [20].

## 5. Outlook

### 5.1 Operational Concept

At the heart of the ESBO concept is the idea of creating a service provider – for instrument developers and for general observers. This concept is common among ground-based, airborne, and also some space telescopes. The telescope hardware and flight platform will be provided and operated by a central organization, or a consortium. Particularly in the short- and mid-term, ESBO will provide flight opportunities for instruments developed by scientific teams to be flown on the telescope ("PI Instruments"), within a shared-time approach. In the long term, turning instruments into facility instruments might be considered as an option.

While part of the available observation time will be reserved, the ESBO concept relies on making a share available to the community via open calls for proposals. A time allocation committee reviews the proposals, and grants observation time.

The approach will be similar to the operating fashion of SOFIA or the European Southern Observatory's (ESO) telescopes. Similar programs also exist for general-purpose balloon experiments, where a balloon gondola is provided by an operator and experiment proposals are invited, such as the French/Canadian "Stratos" program or the DLR/SSC/ESA BEXUS program.

ESBO will extend this approach to astronomical applications by offering: (i) Peer-reviewed, proposal-based access to stratospheric observations for the astronomical community; (ii) Regular and well controlled flight opportunities for astronomical instruments with secured recovery and return of instruments.

### 5.2 Timeline

The plans for ESBO foresee a step-wise development, which is outlined in Figure 5. The current prototype development and conceptual design under ESBO *DS* will be concluded in 2021 and lead to the STUDIO prototype flight shortly thereafter. This flight will simultaneously serve as a first science precursor of the 0.5 m flight system. Further scientific and technology test re-flights of the modified STUDIO payload and gondola are foreseen thereafter. ESBO *DS* will also serve to develop a user group for further payloads, also for a mid-sized flight infrastructure to be potentially added in the 2023/2024 timeframe. Regular operation of the 5 m class FIR flight system is estimated in a 15 years timeframe.

### Acknowledgements

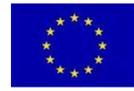ESBO *DS* has received funding from the European Union's Horizon 2020 research and innovation programme under grant agreement No 777516.

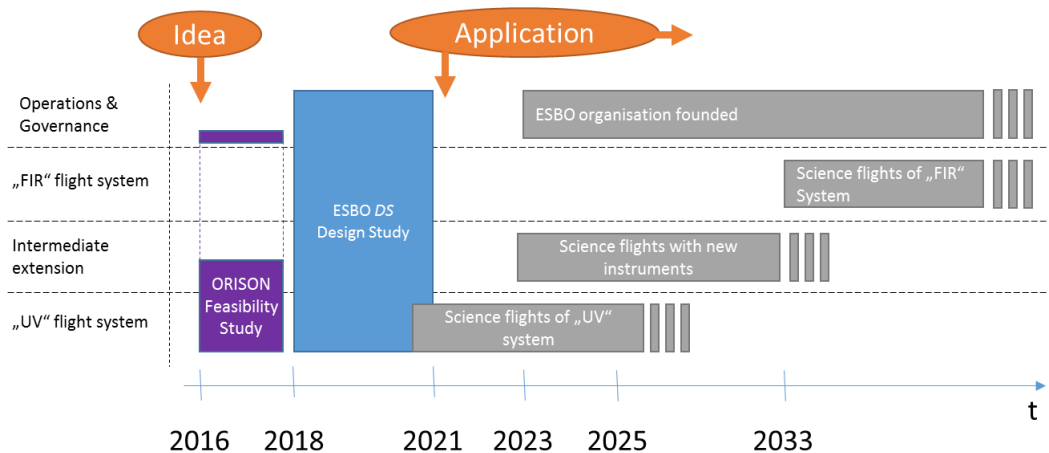

Fig. 5. Preliminary ESBO timeline. All future projections (grey) are preliminary plans based on current estimates and will be refined in a development roadmap during ESBO *DS*.